\documentclass{article}

\usepackage{amsmath,amssymb,amsfonts}
\usepackage{a4wide}
\usepackage{color}
\usepackage{bm}
\usepackage{graphicx}
\usepackage{ulem}

\DeclareMathOperator*{\tr}{tr}

\definecolor{gray}{gray}{0.75}

\newcommand{\1}{\mbox{1}\hspace{-0.25em}\mbox{l}}
\newcommand{\bL}{\bm{L}}
\newcommand{\bT}{\bm{T}}
\newcommand{\bs}{\bm{s}}

\newcommand{\barphi}{\overline{\varphi}}

\newcommand{\remove}[1]{\if0{#1}\fi}  
\newcommand{\comment}[1]{\if0{#1}\fi} 
\newcommand{\add}[1]{{#1}}

\newcommand{\remtomaz}[1]{\if0{#1}\fi}
\newcommand{\addtomaz}[1]{{#1}}

\date{\today}
\title{Construction of the steady state density matrix and quasilocal charges for the spin-$1/2$ XXZ chain with boundary magnetic fields}
\author{Chihiro Matsui$^1$ and Toma\v{z} Prosen$^2$ \\[3ex]
{\it $^1$ Graduate School of Mathematical Sciences, The University of Tokyo} \\
{\it 3-8-1, Komaba, Meguro-ku, 153-8914 Tokyo, Japan} \\
{\it $^2$ Department of Physics, Faculty of Mathematics and Physics,} \\
{\it University of Ljubljana, Jadranska 19, 1000 Ljubljana, Slovenia}}

\begin{document}
\maketitle

\begin{center}
{\bf Abstract}
\end{center}
\bigskip
{\small
We construct the nonequilibrium steady state (NESS) density operator of the spin-$1/2$ XXZ chain with \remove{arbitrary}\add{non-diagonal} boundary magnetic fields \add{coupled to boundary dissipators}. The\remove{corresponding} Markovian boundary dissipation is found with which the NESS density operator is expressed in terms of the product of the Lax operators by relating the dissipation parameters to the boundary parameters of the spin chain. 
The NESS density operator can be expressed in terms of a non-Hermitian transfer operator (NHTO) which forms a commuting family \add{of quasilocal charges}. The optimization of the Mazur bound for the high temperature Drude weight is discussed by using the quasilocal charges and the conventional local charges constructed \remove{in the context of}\add{through} the Bethe ansatz. 
}

\section{Introduction}
The Heisenberg spin chain with anisotropy, the so-called XXZ chain, is one of the most widely studied quantum systems. Its integrability allows us to diagonalize the Hamiltonian~\cite{bib:B31, bib:TF84}, which leads to the derivation of exact physical quantities such as correlation functions~\cite{bib:KBI93, BIB:S89, bib:KMT00, bib:JMMN92}. The integrability comes from the decomposability of many-body scatterings into a sequence of two-body scatterings. The decomposability is guaranteed by the Yang-Baxter equation satisfied by \addtomaz{the} scattering matrices. 

However, it is still a challenging problem to unveil the model's nonequilibrium properties. Peculiar nonequilibrium behaviors of integrable systems result from the existence of sufficiently many conserved quantities 
\remtomaz{to determine}\addtomaz{so that they are in one-to-one correspondence with the} degrees of freedom. For instance, it was predicted that observables after relaxation are described by the generalized Gibbs ensemble (GGE)~\cite{bib:RMO06, bib:RDYO07} by maximizing the entropy, instead of the grand canonical ensemble. Subsequently, it has been demonstrated that integrable systems exhibit generalized thermalization~\cite{bib:RMO06, bib:VR16}. 
Another interesting question is, whether an integrable system \remtomaz{shows}\addtomaz{exhibits} ballistic transport at finite or high temperatures. The optimized lower bound on the ballistic transport coefficient -- the so-called Drude weight -- has been introduced~\cite{bib:ZNP97} by using a commuting family of local charges, although this bound generically vanishes when the system possesses the $\mathbb{Z}_2$-symmetry with respect to which the transporting current is odd.

A new and fruitful approach to the above questions came recently with exact solutions of boundary driven \remtomaz{diffusive}\addtomaz{open} quantum systems~\cite{bib:P11, bib:P11*}. The idea to derive the 
explicit nonequilibrium steady state (NESS) density operator lies in the matrix product ansatz (MPA), which was originally introduced for constructing the ground state of the $s=1$ spin chain~\cite{bib:KSZ93} and widely used to solve classical boundary driven diffusive many-body systems in one-dimension~\cite{bib:DEHP93}. 
The matrix product states serving as the NESS density operator for the quantum boundary driven\remtomaz{diffusive} system is given by the non-Hermitian transfer operator (NHTO) constructed from the product of the Lax operators but with the complex spin representation for the auxiliary space~\cite{bib:P11, bib:P11*}. Due to the Yang-Baxter equation, the NHTO satisfies the divergence condition~\cite{bib:S70, bib:P15}, which implies bulk cancellation in the steady-state Lindblad master equation with the remainder terms localized at the boundaries.
 These terms are in turn compensated by the boundary dissipation, which turns the NHTO into the NESS density operator. 
Amazingly, the NESS density operator constructed in this way can serve, in the limit of small dissipation, as a novel quasilocal conserved quantity, which can be used to evaluate the lower bound for the high temperature Drude weight on the corresponding non-dissipative quantum system~\cite{bib:PI13, bib:IP13}. Indeed, the NESS satisfies both commutativity and quasilocality conditions~\cite{bib:PI13, bib:P14}. 
What made the long-standing problem, {\it i.e.} to find the NESS of the quantum boundary-driven diffusive many-body system, solved is the complex spin representation of the auxiliary space, which has not been considered before. This extension turned out to be important also in the context of integrable systems, since the NHTO provides a new commuting family of conserved quantities which contain \addtomaz{parts that are} orthogonal to the known local conserved quantities. 

We aim in this paper to investigate how boundary magnetic fields imposed on the spin chain affect its nonequilibrium behavior. We derive the NESS density operator of the boundary-driven\remtomaz{diffusive} quantum spin chain with arbitrary boundary magnetic fields. Interestingly, there always exists the corresponding boundary dissipation for which the NESS density operator is given in terms of the NHTO by properly choosing the dissipation rates as functions of boundary fields. 
We have also constructed the quasilocal charges for the corresponding non-dissipative spin chain. We showed that, even under the existence of arbitrary boundary magnetic fields, the NHTO forms a commuting family by keeping quasilocality. These two properties of the NHTO allow us to evaluate the lower bound for the high temperature Drude weight. We found optimization of the optimized Mazur bound due to the $\mathbb{Z}_2$-symmetry breaking in the spin chain with \remove{arbitrary}\add{non-diagonal} boundary magnetic fields, which leads to a finite contribution of the conventional local charges to the lower bound. 

This paper is organized as follows. In the next section, we give the basics of the open spin-$1/2$ XXZ chain and the boundary dissipator of the Lindblad type. We derive the NESS density operator in Section 3. The quasilocal charges are constructed in Section 4. The lower bound for the high temperature Drude weight is also evaluated. The last section is devoted to the concluding remarks. 

\section{The open spin-$1/2$ XXZ chain with boundary dissipation}
\subsection{The spin-$1/2$ XXZ chain with non-diagonal boundaries}
Let us consider the spin-$1/2$ XXZ chain with arbitrary boundary magnetic fields: 
\begin{equation} \label{eq:hamiltonian}
 H = \sum_{x=1}^{n-1} \1_{2^{x-1}} \otimes h \otimes \1_{2^{n-x-1}} + h_{\rm B,L} \otimes \1_{2^{n-1}} + \1_{2^{n-1}} \otimes h_{\rm B,R}, 
\end{equation}
where $\1_{2^x}$ represents the $x$-fold tensor product of the $2\times 2$ identity matrix. The Hamiltonian density for the bulk part $h$ is expressed by the Pauli matrices $\sigma^{\alpha}$ ($\alpha \in \{\pm, z\}$) as 
\begin{equation}
 h = 2 \sigma^+ \otimes \sigma^- + 2 \sigma^- \otimes \sigma^+ + \sigma^z \otimes \sigma^z \cos\eta, 
\end{equation}
while the boundary Hamiltonian density \addtomaz{is expressed} as 
\begin{equation} \label{eq:boundary_h}
\begin{split}
 &h_{\rm B,L} = \frac{1}{2}\sigma^z \sin\eta \cot\xi_{\rm L} + \sigma^+ \kappa_{\rm L} e^{\theta_{\rm L}} \frac{\sin\eta}{\sin\xi_{\rm L}} + \sigma^- \kappa_{\rm L} e^{-\theta_{\rm L}} \frac{\sin\eta}{\sin\xi_{\rm L}}, \\
 &h_{\rm B,R} = \frac{1}{2}\sigma^z \sin\eta \cot\xi_{\rm R} + \sigma^+ \kappa_{\rm R} e^{\theta_{\rm R}} \frac{\sin\eta}{\sin\xi_{\rm R}} + \sigma^- \kappa_{\rm R} e^{-\theta_{\rm R}} \frac{\sin\eta}{\sin\xi_{\rm R}}, 
\end{split}
\end{equation}
containing six free parameters $\xi_{\rm L,R}, \kappa_{\rm L,R},$ and $\theta_{\rm L,R}$ which uniquely parametrise arbitrary boundary magnetic fields. 

The model \eqref{eq:hamiltonian} is known to be integrable in the sense that its transfer matrix forms a commuting family of infinitely many local operators. The local charges are obtained by expanding the logarithm of the transfer matrix around the permutation point. The leading term gives the momentum operator, while the next-to-leading term gives the Hamiltonian: 
\begin{equation}
 H = \frac{d}{d\varphi} 
  \left(\frac{\sin\eta}{2\sin\xi_{\rm L}} K_1\left(\varphi,\xi_{\rm L}\right) + \sum_{x=1}^{n-1} 2\check{R}_{x,x+1}\left(\varphi\right) \right)\Bigg|_{\varphi=\frac{\eta}{2}}
  + \frac{\tr_0 K_0\left(\varphi+\eta,\xi_{\rm R}\right)h_{n,0}}{\tr K_0\left(\varphi+\eta,\xi_{\rm R}\right)} \Bigg|_{\varphi=\frac{\eta}{2}}. 
\end{equation}
The $\check{R}$-matrix and the $K$-matrix satisfy the so-called $RLL$ relation and the reflection relation, respectively: 
\begin{align}
 &\check{R}_{1,2}(\varphi_1 - \varphi_2) \bL_1(\varphi_1) \bL_2(\varphi_2) 
 = \bL_1(\varphi_2) \bL_2(\varphi_1) \check{R}_{1,2}(\varphi_1 - \varphi_2), \label{eq:RLL_rel} \\
 &\check{R}_{2,1}(\varphi_1-\varphi_2) K_2(\varphi_1) \check{R}_{1,2}(\varphi_1+\varphi_2) K_2(\varphi_2)
  = K_2(\varphi_2) \check{R}_{2,1}(\varphi_1+\varphi_2) K_2(\varphi_1) \check{R}_{1,2}(\varphi_1-\varphi_2), \label{eq:ref_rel}
\end{align}
whose solutions are expressed \addtomaz{in terms}\remtomaz{by means} of the Pauli matrices $\sigma^{\alpha}$ $(\alpha \in +,-,z)$:  
\begin{align}
 &\check{R}(\varphi) = \frac{\sin\varphi}{2} (h + \1\cos\eta) - \frac{1 + \cos\varphi}{2} \1\sin\eta + \frac{1 - \cos\varphi}{2} \sigma^z \otimes \sigma^z \sin\eta, \\
 &K\left(\varphi;\xi,\kappa,\theta\right) = \1 \sin\xi \cos\varphi + \sigma^z \cos\xi \sin\varphi + \sigma^+ \kappa e^{\theta} \sin(2\varphi) + \sigma^- \kappa e^{-\theta} \sin(2\varphi). 
\end{align}
On the other hand, the Lax operator $L$ consists of the physical space, which has the same representation as that of the $\check{R}$-matrix, and the auxiliary space, which admits any representation including the complex spin representation: 
\begin{equation} 
 \bL(\varphi,s) = 
  \begin{pmatrix}
   \sin(\varphi + \eta\bs_a^z) & \sin\eta \bs_a^- \\
   \sin\eta \bs_a^+ & \sin(\varphi - \eta\bs_a^z)
  \end{pmatrix}
  = \sum_{\alpha \in \{+,-,0,z\}} \bL_a^{\alpha}(\varphi,s) \otimes \sigma_p^{\alpha}, 
\end{equation}
where 
\begin{equation}
 \bL_a^0(\varphi,s) = \sin\varphi \cos(\eta\bs_a^z), \quad
  \bL_a^z(\varphi,s) = \cos\varphi \sin(\eta\bs_a^z), \quad
  \bL_a^{\pm}(\varphi,s) = (\sin\eta) \bs_a^{\mp}. 
\end{equation}
We take the complex spin representations in the way introduced in Ref.~\cite{bib:P15}: 
\begin{equation} \label{eq:spin_rep_nonsym}
\begin{split}
 &\bs_a^z = \sum_{k=0}^{\infty} (s-k) |k \rangle\langle k|, \\
 &\bs_a^+ = \sum_{k=0}^{\infty} \frac{\sin(k+1)\eta}{\sin\eta} |k \rangle\langle k+1|, \\
 &\bs_a^- = \sum_{k=0}^{\infty} \frac{\sin(2s-k)\eta}{\sin\eta} |k+1 \rangle\langle k| 
\end{split}
\end{equation}
by which the $U_q(sl_2)$ algebraic relations are satisfied: 
\begin{equation}
 [\bs^z,\,\bs^{\pm}] = \pm\bs^{\pm}, \quad
  [\bs^+,\,\bs^-] = [2\bs^z]_q. 
\end{equation}

\subsection{Boundary dissipation of the Lindblad type}
Within the theory of open quantum systems~\cite{bib:BP02}, incoherent Markovian quantum dissipation is completely described by a set of Lindblad operators $\{L_{\mu} \in {\rm End}(\mathcal{H}_p^{\otimes n}),\, \mu=1,2,\dots\}$. Such a system's many-body density operator $\rho(t)$ obeys the time evolution described by the Lindblad-Gorini-Kossakowski-Sudarshan master equation~\cite{bib:GKS76, bib:L76}: 
\begin{equation} \label{eq:LB1}
 \frac{d}{dt} \rho(t) = \hat{\mathcal{L}} \rho(t)
  := -i[H,\, \rho(t)] + \sum_{\mu} \Big( 2L_{\mu} \rho(t) L_{\mu}^{\dag} - \{L_{\mu}^{\dag} L_{\mu},\, \rho(t)\} \Big). 
\end{equation}

Throughout this paper, we consider the ultra-local Lindblad operators 
\begin{equation}
 L_{\mu} = \ell_{\mu} \otimes \1_{2^{n-1}} \quad
  {\rm or} \quad
  L_{\mu} = \1_{2^{n-1}} \otimes \ell_{\mu}, \quad
  \ell_{\mu} \in {\rm End} (\mathcal{H}_p), 
\end{equation}
especially of the following forms associated with three different dissipation rates $\varepsilon, \varepsilon', \varepsilon'' \in \mathbb{R}^+$ and two additional dissipaiton parameters $\alpha,\alpha'\in\mathbb{C}$:
\begin{equation}
 L_1 = \sqrt{\varepsilon} (\sigma_1^+ + \alpha\sigma_1^0), \quad
  L_2 = \sqrt{\varepsilon} (\sigma_n^- + \alpha'\sigma_n^0), \quad
  L_3 = \sqrt{\varepsilon'} \sigma_1^z, \quad
  L_4 = \sqrt{\varepsilon''} \sigma_n^z. 
\end{equation}
\remtomaz{One clearly obtains}\addtomaz{Notice} that the dissipators $L_1$ and $L_3$ are coupled to the left boundary of the spin chain, while $L_2$ and $L_4$ to the right boundary. We let $\alpha$ and $\alpha'$ free for the moment and determine their relations to the boundary parameters later. 

The Liouvillian $\hat{\mathcal{L}}$ is then written as 
\begin{equation} \label{eq:Liouvillian}
 \hat{\mathcal{L}} = -i [H,\,\rho(t)] + \varepsilon \hat{\mathcal{D}}_{\sigma_1^+ + \alpha\sigma_1^0} + \varepsilon' \hat{\mathcal{D}}_{\sigma_1^z} + \varepsilon \hat{D}_{\sigma_n^- + \alpha'\sigma_n^0} + \varepsilon'' \hat{\mathcal{D}}_{\sigma_n^z}, 
\end{equation}
where the dissipator map is defined by 
\begin{equation}
 \hat{\mathcal{D}}_L (\rho) = 2L \rho L^{\dag} - \{L^{\dag} L,\,\rho\}. 
\end{equation}

From the master equation \eqref{eq:LB1}, the density operator at time $t$ is expressed as $\rho(t) = \exp(t \hat{\mathcal{L}}) \rho(0)$ leading to, if the limit exists, the NESS density operator $\rho_{\infty} = \lim_{t \to \infty} \exp(t \hat{\mathcal{L}}) \rho(0)$. Since the NESS is invariant under the time development, the NESS density operator $\rho_{\infty}$ gives the fixed point of the propagator: 
\begin{equation} \label{eq:fixed_p}
 \hat{\mathcal{L}} \rho_{\infty} = 0. 
\end{equation}

\section{Nonequilibrium steady state}
\subsection{Construction of the NESS density operator}
Let us first write the NESS density operator in terms of the product of non-Hermitian amplitude operators: 
\begin{equation} \label{eq:NESS}
 \rho_{\infty} = \frac{R_{\infty}}{\tr R_{\infty}}, \quad
  R_{\infty} = \Omega_n \Omega_n^{\dag}. 
\end{equation}
Contrary to the trivial open boundary case where a particular factorization occurs in terms of $\Omega_n$~\cite{bib:P15}, some modification is required \remove{for the arbitrary boundary case}\add{under the presence of boundary magnetic fields}. 

The fixed point condition \eqref{eq:fixed_p} for the NESS density operator $\rho_{\infty}$ under the choice of the Liouvillian \eqref{eq:Liouvillian} leads to the following condition on $\Omega_n$ and $\Omega_n^{\dag}$: 
\begin{equation} \label{eq:fixed_p_cond}
 i\varepsilon^{-1} [H,\,\Omega_n \Omega_n^{\dag}] 
  = \hat{\mathcal{D}}_{\sigma_1^+ + \alpha\sigma_1^0} (\Omega_n \Omega_n^{\dag}) + \frac{\varepsilon'}{\varepsilon} \hat{\mathcal{D}}_{\sigma_1^z} (\Omega_n \Omega_n^{\dag}) + \hat{\mathcal{D}}_{\sigma_n^- + \alpha'\sigma_n^0} (\Omega_n \Omega_n^{\dag}) + \frac{\varepsilon''}{\varepsilon} \hat{\mathcal{D}}_{\sigma_n^z} (\Omega_n \Omega_n^{\dag}). 
\end{equation}
In order to deal with the product of the amplitude operators, it is useful to introduce the double Lax operator \addtomaz{acting over a tensor product of a pair of complex spin representations}~\cite{bib:P15}.\remtomaz{We use the Lax operator with the complex spin representation for the auxiliary space.} Note that the complex spin operators have the highest weight representations given in \eqref{eq:spin_rep_nonsym}. Besides, we introduce another complex spin operators of the transposed lowest weight representations: 
\begin{equation}
\begin{split}
 &t^z = \sum_{k=0}^{\infty} (t-k) |k \rangle \langle k|, \\
 &t^+ = \sum_{k=0}^{\infty} \frac{\sin(k+1)\eta}{\sin\eta} |k+1 \rangle \langle k|, \\
 &t^- = \sum_{k=0}^{\infty} \frac{\sin(2t-k)\eta}{\sin\eta} |k \rangle \langle k+1|. 
\end{split}
\end{equation}
The double Lax operators is then defined in the product representation $\mathcal{V}_s^T \otimes \mathcal{V}_t$: 
\begin{equation} \label{eq:double_L}
 \mathbb{L}_x(\varphi,\theta,s,t) = \bL_{a,x}^T(\varphi,s) \bL_{b,x}(\theta,t), 
\end{equation}
where $\mathcal{V}_s^T$ is the transposed lowest weight representation, while $\mathcal{V}_t$ is the highest weight representation. 

The double Lax operator obeys the Yang-Baxter-like equation: 
\begin{equation}
\begin{split}
 &\check{R}_{1,2}(\delta_1-\delta_2) \mathbb{L}_1(\varphi+\delta_1, \theta-\delta_1, s, t) \mathbb{L}_2(\varphi+\delta_2, \theta-\delta_2, s, t) \\
 &= \mathbb{L}_1(\varphi+\delta_2, \theta-\delta_2, s, t) \mathbb{L}_2(\varphi+\delta_1, \theta-\delta_1, s, t) \check{R}_{1,2}(\delta_1-\delta_2), 
\end{split}
\end{equation}
from which we obtain the \remove{divergence}\add{commutativity} condition for the bulk part: 
\begin{equation} \label{eq:div_bulk}
 [H_{\rm bulk},\,\mathbb{L}_1 \cdots \mathbb{L}_n]
  = 2\sin\eta ( \partial\mathbb{L}_1 \mathbb{L}_2 \cdots \mathbb{L}_n - \mathbb{L}_1 \cdots \mathbb{L}_{n-1} \partial\mathbb{L}_n), 
\end{equation}
where 
\begin{equation}
\begin{split}
 \partial\mathbb{L}_x(\varphi, \theta, s, t) 
 &= \partial_{\delta} \Big( \bL_{a,x}^T(\varphi+\delta, s) \bL_{b,x}(\theta-\delta, t) \Big)_{\delta=0} \\
 &= \partial_{\varphi}\bL_{a,x}^T(\varphi,s) \bL_{b,x}(\theta,t) - \bL_{a,x}^T(\varphi,s) \partial_{\theta}\bL_{b,x}(\theta,t). 
\end{split}
\end{equation}
The bulk \remove{divergence}\add{commutativity} condition implies that the dissipation should be localized at the boundaries of the spin chain. 
Using the bulk \remove{divergence}\add{commutativity} condition \eqref{eq:div_bulk} and the boundary Hamiltonian density \eqref{eq:boundary_h}, we obtain the \remove{divergence}\add{commutativity} condition for the full spin chain: 
\begin{equation} \label{eq:div_full}
\begin{split}
 [H,\,\mathbb{L}^{\otimes n}]
 &= [H_{\rm bulk},\,\mathbb{L}^{\otimes n}] + [H_{\rm L,B},\,\mathbb{L}^{\otimes n}] + [H_{\rm R,B},\,\mathbb{L}^{\otimes n}] \\
 &= 2(\sin\eta) \partial\mathbb{L} \otimes \mathbb{L}^{\otimes n-1} - 2(\sin\eta) \mathbb{L}^{\otimes n-1} \otimes \partial\mathbb{L} \\
 &+ \Big[ \frac{1}{2}\sigma^z \sin\eta \coth\xi_{\rm L} + \sigma^+ \kappa_{\rm L} e^{\theta_{\rm L}} \frac{\sin\eta}{\sin\xi_{\rm L}} + \sigma^- \kappa_{\rm L} e^{-\theta_{\rm L}} \frac{\sin\eta}{\sin\xi_{\rm L}},\,\mathbb{L} \Big] \otimes \mathbb{L}^{\otimes n-1} \\
 &+ \mathbb{L}^{\otimes n-1} \otimes \Big[ \frac{1}{2}\sigma^z \sin\eta \coth\xi_{\rm R} + \sigma^+ \kappa_{\rm R} e^{\theta_{\rm R}} \frac{\sin\eta}{\sin\xi_{\rm R}} + \sigma^- \kappa_{\rm R} e^{-\theta_{\rm R}} \frac{\sin\eta}{\sin\xi_{\rm R}},\,\mathbb{L} \Big]. 
\end{split}
\end{equation}

We assume that the NESS density operator is given by the product of the Lax operators, similarly as for the trivial open boundary case~\cite{bib:P14}: 
\begin{equation}
 \Omega_n = \frac{1}{\sin^n(\varphi + s\eta)} 
  {_a}\langle 0| \bL_1^T \bL_2^T \dots \bL_n^T |0 \rangle_a, 
\end{equation}
where $|0 \rangle$ is the highest weight vector. Note that we have used the partially transposed Lax operator\remove{$\mathbb{L}_x \in \mathcal{V}_s^T \otimes \mathcal{V}_{\bar{s}}$}: 
\begin{equation}
 \bL^T(\varphi,s) = \sum_{\alpha \in \{+,-,0,z\}} \bL^{\alpha}(\varphi,s) \otimes (\sigma^{\alpha})^T. 
\end{equation}
From the definition~\eqref{eq:double_L}, the product $\Omega_n \Omega_n^{\dag}$ admits the expression in terms of the double Lax operator:  
\begin{equation}
 \Omega_n \Omega_n^{\dag} = \frac{1}{\sin^n(\varphi + s\eta) \sin^n(\varphi + \bar{s}\eta)}
  {_a}\langle 0| {_b}\langle 0| \mathbb{L}_1 \cdots \mathbb{L}_n |0 \rangle_a |0 \rangle_b. 
\end{equation}
By applying the double highest weight vector $|0 \rangle_a |0 \rangle_b$ to the fixed point condition \eqref{eq:fixed_p_cond}, we obtain the left and right boundary conditions: 
\begin{equation} \label{eq:boundary_cond*}
\begin{split}
 &{_a}\langle 0| {_b}\langle 0| \Big( -i\varepsilon^{-1} 2(\sin\eta) \partial\mathbb{L} 
 - i\varepsilon^{-1} \Big[ \frac{1}{2}\sigma^z \sin\eta \coth\xi_{\rm L} + \sigma^+ \kappa_{\rm L} e^{\theta_{\rm L}} \frac{\sin\eta}{\sin\xi_{\rm L}} + \sigma^- \kappa_{\rm L} e^{-\theta_{\rm L}} \frac{\sin\eta}{\sin\xi_{\rm L}},\,\mathbb{L} \Big] \\
 &\hspace{96mm}+ \hat{D}_{\sigma_1^+ + \alpha\sigma_1^0}(\mathbb{L}) + \frac{\varepsilon'}{\varepsilon} \hat{D}_{\sigma_1^z}(\mathbb{L}) \Big) = 0, 
 \\
 &\Big( i\varepsilon^{-1} 2(\sin\eta) \partial\mathbb{L} 
 - i\varepsilon^{-1} \Big[ \frac{1}{2}\sigma^z \sin\eta \coth\xi_{\rm R} + \sigma^+ \kappa_{\rm R} e^{\theta_{\rm R}} \frac{\sin\eta}{\sin\xi_{\rm R}} + \sigma^- \kappa_{\rm R} e^{-\theta_{\rm R}} \frac{\sin\eta}{\sin\xi_{\rm R}},\,\mathbb{L} \Big] \\
 &\hspace{83mm}+ \hat{D}_{\sigma_n^- + \alpha'\sigma_n^0}(\mathbb{L}) + \frac{\varepsilon''}{\varepsilon}\hat{D}_{\sigma_n^z}(\mathbb{L}) \Big) |0 \rangle_a |0 \rangle_b = 0. 
\end{split}
\end{equation}
Note that the dissipation operators act on a\remtomaz{n} $2$-by-$2$ matrix as 
\begin{align}
 &\hat{D}_{\sigma^+ + \alpha\sigma^0}
 \begin{pmatrix} a & b \\ c & d \end{pmatrix}
 = \begin{pmatrix}  
    2d + \alpha(b+c) & -b - \alpha(a-d) \\ -c - \alpha(a-d) & -2d - \alpha(b+c)
   \end{pmatrix}, \\
 &\hat{D}_{\sigma^- + \alpha\sigma^0}
 \begin{pmatrix} a & b \\ c & d \end{pmatrix}
 = \begin{pmatrix}
    -2a - \alpha(b+c) & -b + \alpha(a-d) \\ -c + \alpha(a-d) & 2a + \alpha(b+c)
   \end{pmatrix}, \\
 &\hat{D}_{\sigma^z}
 \begin{pmatrix} a & b \\ c & d \end{pmatrix}
 = \begin{pmatrix}
    0 & -4b \\ -4c & 0
   \end{pmatrix}. 
\end{align}

The nontrivial solution to \eqref{eq:boundary_cond*} exists for the following case. We first need the spectral parameter restricted by 
\begin{equation}
 \varphi = \frac{\pi}{2}. 
\end{equation}
This condition coincides with the trivial open boundary case~\cite{bib:P15}. The dissipation rates must be related to the anisotropy and the boundary parameters of the diagonal parts: 
\begin{equation}
 \varepsilon = -2i \sin\eta \tan(s\eta), \quad
  \varepsilon' = -\frac{i}{4} \sin\eta \cot\xi_{\rm L}, \quad
  \varepsilon'' = -\frac{i}{4} \sin\eta \cot\xi_{\rm R}. 
\end{equation}
This result includes the trivial open boundary case for $\xi_{\rm L,R} = \pi/2$~\cite{bib:P14}. The parameters $\alpha$ and $\alpha'$ are determined by the boundary parameters of the non-diagonal parts: 
\begin{equation}
 \alpha = i\varepsilon^{-1} \frac{\sin\eta}{\sin\xi_{\rm L}} \kappa_{\rm L} e^{\theta_{\rm L}}
  = -i\varepsilon^{-1} \frac{\sin\eta}{\sin\xi_{\rm L}} \kappa_{\rm L} e^{-\theta_{\rm L}}, \quad
  \alpha' = -i\varepsilon^{-1} \frac{\sin\eta}{\sin\xi_{\rm R}} \kappa_{\rm R} e^{\theta_{\rm R}}
  = i\varepsilon^{-1} \frac{\sin\eta}{\sin\xi_{\rm R}} \kappa_{\rm R} e^{-\theta_{\rm R}}, 
\end{equation}
which require $\theta_{\rm L} = \theta_{\rm R} = i\pi/2$. As a result, the boundary Hamiltonian density is allowed to take only the following form: 
\begin{align}
 &h_{\rm B,L} = \frac{1}{2}\sigma^z \sin\eta \cot\xi_{\rm L} - 2 \sigma^y \kappa_{\rm L} \frac{\sin\eta}{\sin\xi_{\rm L}}, \\
 &h_{\rm B,R} = \frac{1}{2}\sigma^z \sin\eta \cot\xi_{\rm R} - 2 \sigma^y \kappa_{\rm R} \frac{\sin\eta}{\sin\xi_{\rm R}}. 
\end{align}

\subsection{Rotation in the $xy$-plane}
One may wonder why the solution to the boundary conditions requires the non-diagonal parts of the left and right boundary Hamiltonian to consist of only $\sigma^y$-terms, although the Hamiltonian itself possesses the symmetry with respect to the rotation in the $xy$-plane. 
Let us consider the transformation performed by the matrix defined by 
\begin{equation}
 U := \begin{pmatrix} e^{i\frac{\phi}{2}} & 0 \\ 0 & e^{-i\frac{\phi}{2}} \end{pmatrix}. 
\end{equation}
The transformation makes $\sigma^z$ invariant $U \sigma^z U^{-1} = \sigma^z$ but rotates the other components of the Pauli matrices: 
\begin{align}
 &U \sigma^x U^{-1} = \sigma^x \cos\phi + \sigma^y \sin\phi, \\
 &U \sigma^y U^{-1} = \sigma^x \sin\phi - \sigma^y \cos\phi, 
\end{align}
and, subsequently, 
\begin{equation}
 U \sigma^+ U^{-1} = e^{i\phi} \sigma^+, \quad
  U \sigma^- U^{-1} = e^{-i\phi} \sigma^-. 
\end{equation}
Therefore, it is \remtomaz{easily obtained}\addtomaz{easy to show that} the bulk part of the Hamiltonian \addtomaz{is} invariant under the transformation $U^{\otimes n} H_{\rm bulk} (U^{\otimes n})^{-1} = H_{\rm bulk}$. Also, the $\sigma^z$-terms of the left and \addtomaz{the} right boundary Hamiltonian are invariant under the transformation. Thus, $U^{\otimes n}$ rotates the Hamiltonian in the $xy$-plane by letting the non-diagonal boundary terms contain both $\sigma^x$- and $\sigma^y$-terms. 

Let us denote the transformed operator by $\widetilde{X} := U X U^{-1}$. The boundary Hamiltonian density is then deformed as 
\begin{align}
 &h_{\rm B,L} \to
 \widetilde{h}_{\rm B,L} = \frac{1}{2}\sigma^z \sin\eta \cot\xi_{\rm L} - 2 \Big(\sigma^x \sin\phi - \sigma^y \cos\phi\Big) \kappa_{\rm L} \frac{\sin\eta}{\sin\xi_{\rm L}}, \\
 &h_{\rm B,R} \to
 \widetilde{h}_{\rm B,R} = \frac{1}{2}\sigma^z \sin\eta \cot\xi_{\rm R} - 2 \Big(\sigma^x \sin\phi - \sigma^y \cos\phi\Big) \kappa_{\rm R} \frac{\sin\eta}{\sin\xi_{\rm R}}.
\end{align}
The transformation angle $\phi$ is related to the non-diagonal boundary parameters as 
\begin{equation}
 i\theta_{\rm L,R} = \phi + \frac{\pi}{2}, 
\end{equation}
which implies that the only condition for obtaining the NESS density operator is $\theta_{\rm L} = \theta_{\rm R}$. 
Indeed, the fixed point condition \eqref{eq:fixed_p_cond} holds for the transformed Hamiltonian $\widetilde{H}$ and double Lax operator $\widetilde{\mathbb{L}}$ if we deform the Lindblad operators as 
\begin{equation}
\begin{split}
 &\hat{D}_{\sigma^+ + \alpha\sigma^0} \to \hat{D}_{\widetilde{\sigma}^+ + \alpha\widetilde{\sigma}^0} = \hat{D}_{e^{i\phi}\sigma^+ + \alpha\sigma^0}, \\
 &\hat{D}_{\sigma^- + \alpha\sigma^0} \to \hat{D}_{\widetilde{\sigma}^- + \alpha\widetilde{\sigma}^0} = \hat{D}_{e^{-i\phi}\sigma^- + \alpha\sigma^0}, \\
 &\hat{D}_{\sigma^z} \to \hat{D}_{\widetilde{\sigma}^z} = \hat{D}_{\sigma^z}. 
\end{split}
\end{equation}

\section{Quasilocal charges}
\subsection{Optimized Mazur bound on the  Drude weight}
The ballistic \addtomaz{transport} is characterized by the finite Drude weight $D$. The Drude weight is defined as the coefficient for the diverging part of d.c. conductivity in the context of linear-response transport. Since the Drude weight has the temperature dependence, its expansion at high temperature leads to 
\begin{equation} \label{eq:Drude}
 D = \beta D_{\infty} + \mathcal{O}(\beta^2), \quad
  D_{\infty} = \lim_{t \to \infty} \lim_{n \to \infty}
  \frac{1}{2tn} \int_0^t dt' (J(t'), J), 
\end{equation}
where $J$ is the extensive current $J := \sum_{x=0}^{n-1} \1_{2^{x}} \otimes j \otimes \1_{2^{n-x}}$, \addtomaz{$j$ is the local current,} and $(A,B)$ is the Hilbert-Schmidt inner product defined by $(A,B) := \tr (A^{\dag}B)/\tr \1_{2^n}$. 

By using a set of \addtomaz{extensive local} conserved operators $\{Q_r;\,r=1,\dots,n\}$, the Mazur bound~\cite{bib:M69} allows rigorous estimation of the high-temperature Drude weight from below~\cite{bib:ZNP97}: 
\begin{equation} \label{eq:Drude_high}
\begin{split}
 &D_{\infty} \geq \lim_{n \to \infty} \frac{1}{2n} \sum_{r,r'} (J,Q_r) (K^{-1})_{r,r'} (Q_{r'},J) 
 := D_Q, 
\\
 &K_{r,r'} := (Q_r,Q_{r'}).
\end{split}
\end{equation}
However, in the system with the $\mathbb{Z}_2$-symmetry such as the spin-reversal symmetry, \addtomaz{with respect to which the conserved charges $Q_r$ are even and the current $J$ is odd}, 
the lower bound $D_Q$ always becomes zero, which tells nothing about the ballistic behavior. 

The situation drastically changes by the introduction of quasilocal charges~\cite{bib:PI13, bib:P14}. The quasilocal charges constructed by differentiating the NESS density operator for the corresponding boundary driven\remtomaz{diffusive} spin chain with respect to the representation parameter $s$ consists of both even and odd parity parts, which makes the lower bound finite. The optimized Mazur bound~\cite{bib:P14, bib:P15} was obtained as 
\begin{equation} \label{eq:Mazur_b}
 D_{\infty} \geq \frac{1}{2} {\rm Re} \int_{\mathcal{D}_m} d^2\varphi Z_J(\varphi) f(\varphi) 
  := D_Z, 
\end{equation}
which is bounded by the conventional Drude weight $D_Z \geq D_Q$. Here we used the notation $Z_J(\varphi) := \lim_{n \to \infty} (J,Z(\varphi))/n$, \addtomaz{and $Z(\varphi)$ is the family of quasilocal charges which can be generated from NESS density operator of the boundary driven chain in the limit of weak driving.} The function $f(\varphi)$ is obtained as the solution of a complex Fredholm equation of the first kind: 
\begin{equation} \label{eq:variation}
 \int_{\mathcal{D}_m} d^2\varphi'\, \lim_{n \to \infty} \frac{1}{n} (Z(\varphi),Z(\varphi')) f(\varphi')
  = \overline{Z_J(\varphi)},
  \quad
 \varphi \in \mathcal{D}_m \subset \mathbb{C}. 
\end{equation}

\subsection{Construction of quasilocal charges}
Thus, what we need for evaluating the lower bound of the high temperature Drude weight is to find a commuting family of quasilocal charges. Analogously to the periodic and trivial open boundary cases~\cite{bib:P14, bib:P15}, let us introduce the NHTO defined by the product of the Lax operators: 
\begin{equation}
 W_n(\varphi,s) = \langle 0| \bL(\varphi,s)^{\otimes_pn} |0 \rangle. 
\end{equation}
Note that the commutativity holds for the NHTO with any pair of spectral and representation parameters in spite of the existence of \remove{arbitrary}\add{non-diagonal} boundary magnetic fields: 
\begin{equation}
\begin{split}
 W_n(\varphi,s) W_n(\theta,t) 
 &= \langle 0|_a \langle 0|_b \check{R}_{a,b}(\varphi-\theta,s,t) \Big( \prod_{x=1}^n \bL_{a,x}(\varphi,s) \bL_{b,x}(\theta,t) \Big) |0 \rangle_a |0 \rangle_b \\
 &= \langle 0|_a \langle 0|_b \Big( \prod_{x=1}^n \bL_{a,x}(\varphi,s) \bL_{b,x}(\theta,t) \Big) \check{R}_{a,b}(\varphi-\theta,s,t) |0 \rangle_a |0 \rangle_b \\
 &= W_n(\theta,t) W_n(\varphi,s). 
\end{split}
\end{equation}
The NHTO satisfies the following commutativity condition: 
\begin{equation} \label{eq:div_cond2}
\begin{split}
 [H,\,W_n(\varphi,s)] 
 &= -2\sin\eta \Big( \cos\varphi \cos(s\eta) \sigma^0 \otimes W_{n-1}(\varphi,s) - \sin\varphi \sin(s\eta) \sigma^z \otimes W_{n-1}(\varphi,s) \\
 &\hspace{18mm}- \cos\varphi \cos(s\eta) W_{n-1}(\varphi,s) \otimes \sigma^0 + \sin\varphi \sin(s\eta) W_{n-1}(\varphi,s) \otimes \sigma^z \Big)
 \\
 &+ \frac{\sin\eta}{\sin\xi_{\rm L}} \Big( -2\kappa_{\rm L} e^{\theta_{\rm L}} \cos\varphi \sin(s\eta) \sigma^+ \otimes W_{n-1}(\varphi,s) + 2\kappa_{\rm L} e^{-\theta_{\rm L}} \cos\varphi \sin(s\eta) \sigma^- \otimes W_{n-1}(\varphi,s) \Big)
 \\
 &- \frac{\sin\eta}{\sin\xi_{\rm R}} \Big( -2\kappa_{\rm R} e^{-\theta_{\rm R}} \cos\varphi \sin(s\eta) W_{n-1}(\varphi,s) \otimes \sigma^- + 2\kappa_{\rm R} e^{\theta_{\rm R}} \cos\varphi \sin(s\eta) W_{n-1}(\varphi,s) \otimes \sigma^+ \Big)
 \\
 &+ \frac{\sin\eta}{\sin\xi_{\rm L}} \Big( -\cos\xi_{\rm L} (\sin\eta) \sigma^- \otimes W_{n-1}^+(\varphi,s) + \kappa_{\rm L} e^{\theta_{\rm L}} (\sin\eta) \sigma^z \otimes W_{n-1}^+(\varphi,s) \Big)
 \\
 &- \frac{\sin\eta}{\sin\xi_{\rm R}} \Big( -\cos\xi_{\rm R} \sin(2s\eta) W_{n-1}^-(\varphi,s) \otimes \sigma^+ + \kappa_{\rm R} e^{-\theta_{\rm R}} \sin(2s\eta) W_{n-1}^-(\varphi,s) \otimes \sigma^z \Big). 
\end{split}
\end{equation}
Here we introduced 
\begin{equation}
\begin{split}
 W_n^+(\varphi,s) = \langle 1| \bL_1(\varphi,s) \cdots \bL_n(\varphi,s) |0 \rangle, \quad
 W_n^-(\varphi,s) = \langle 0| \bL_1(\varphi,s) \cdots \bL_n(\varphi,s) |1 \rangle. 
\end{split}
\end{equation}

From now on, we show that the differentiation of the NHTO with respect to the representation parameter $s$ forms a family of quasilocal charges. Precisely, the quasilocal charge is given by 
\begin{equation} \label{eq:quasilocal_c*}
 Z_n(\varphi) = \frac{1}{2(\sin\varphi)^{n-2} \eta \sin\eta} \partial_s W_n(\varphi,s)\Big|_{s=0} - \frac{\sin\varphi \cos\varphi}{2\sin\eta} M_n^z, 
\end{equation}
where $M_n^z = \sum_{x=0}^{n-1} \1_{2^x} \otimes \sigma^z \otimes \1_{2^{n-1-x}}$. 
By using the time derivative of NHTO \eqref{eq:div_cond2}, we have 
\begin{equation} \label{eq:div_simple}
\begin{split}
 [H,\,Z_n(\varphi)] 
 &= 2\sin\eta \cot\varphi \Big( -\sigma^0 \otimes Z_{n-1}(\varphi) + Z_{n-1}(\varphi) \otimes \sigma^0 \Big)
 \\
 &+ \sigma^z \otimes \1_{2^{n-1}} - \1_{2^{n-1}} \otimes \sigma^z
 \\
 &+ \frac{1}{2\eta (\sin\varphi)^{n-2} \sin\xi_{\rm L}} 
 \Big( -\cos\xi_{\rm L} (\sin\eta) \sigma^- \otimes \partial_s W_{n-1}^+(\varphi,s)\Big|_{s=0} + \kappa_{\rm L} e^{\theta_{\rm L}} (\sin\eta) \sigma^z \otimes \partial_s W_{n-1}^+(\varphi,s)\Big|_{s=0} \Big)
 \\
 &- \frac{1}{2\eta (\sin\varphi)^{n-2} \sin\xi_{\rm R}}
 \Big( -2\eta \cos\xi_{\rm R} W_{n-1}^-(\varphi,0) \otimes \sigma^+ + 2\eta \kappa_{\rm R} e^{-\theta_{\rm R}} W_{n-1}^-(\varphi,0) \otimes \sigma^z \Big). 
\end{split}
\end{equation}
The third and fourth lines are evaluated through 
\begin{equation} \label{eq:W+}
\begin{split}
 \partial_s W_{n}^+(\varphi,s)\Big|_{s=0} 
 &= \sum_{k=1}^n \langle 1| \bL_1(\varphi,s) \cdots \partial_s \bL_k(\varphi,s) \cdots \bL_n(\varphi,s) |1 \rangle \Big|_{s=0}
 \\
 &= \sum_{k=1}^n (\sin\varphi)^{n-k} \sum_{\alpha_j \in \{+,-,0,z\}} 
 \langle 1| \bL_1^{\alpha_1}(\varphi,s) \dots (\partial_s \bL_k(\varphi,s))^{\alpha_k} |0 \rangle \Big|_{s=0}
 \sigma^{\alpha_1} \otimes \cdots \otimes \sigma^{\alpha_k} \otimes \1_{2^{n-k}}
 \\
 &= \sum_{k=1}^n 2\eta (\sin\varphi)^{n-k} \sum_{\alpha_j \in \{+,-,0,z\}} 
 \langle 1| \bL_1^{\alpha_1}(\varphi,0) \dots \bL^{\alpha_{k-1}}_{k-1}(\varphi,0) |1 \rangle 
 \sigma^{\alpha_1} \otimes \cdots \otimes \sigma^{\alpha_{k-1}} \otimes \sigma^+ \otimes \1_{2^{n-k}}, 
\end{split}
\end{equation}
\begin{equation} \label{eq:W-}
\begin{split}
 W_n^-(\varphi,0) &= \langle 0| \bL_1(\varphi,0) \cdots \bL_n(\varphi,0) |1 \rangle
 \\
 &= \sum_{\alpha_j \in \mathcal{J}} \sin\varphi \langle 0| \bL_2^{\alpha_2}(\varphi,0) \cdots \bL_n^{\alpha_n}(\varphi,0) |1 \rangle
 \sigma^0 \otimes \sigma^{\alpha_2} \otimes \cdots \otimes \sigma^{\alpha_n} \\
 &\hspace{5mm}+ \sum_{\alpha_j \in \{+,-,0,z\}} \sin\eta \langle 1| \bL_2^{\alpha_2}(\varphi,0) \cdots \bL_n^{\alpha_n}(\varphi,0) |1 \rangle
 \sigma_1^- \otimes \sigma_2^{\alpha_2} \otimes \cdots \otimes \sigma_n^{\alpha_n}
 \\
 &= \sum_{k=1}^n \sin\eta (\sin\varphi)^{k-1} 
 \sum_{\alpha_j \in \{+,-,0,z\}} \langle 1| \bL_k^{\alpha_k}(\varphi,0) \cdots \bL_n^{\alpha_n}(\varphi,0) |1 \rangle
 \1_{2^{k-1}} \otimes \sigma^- \otimes \sigma^{\alpha_{k+1}} \otimes \cdots \otimes \sigma^{\alpha_n},  
\end{split}
\end{equation}
by using 
\begin{equation}
\begin{split}
 &\bL_j(\varphi,0) |0 \rangle = \sigma_j^0 \sin\varphi |0 \rangle, \quad
 \partial_s \bL_j(\varphi,s)|_{s=0} |0 \rangle = 2\eta \sigma_j^+ |1 \rangle, \\
 &\langle 0| \bL_j(\varphi,0) = \sigma_j^0 \sin\varphi \langle 0| + \sigma_j^- \sin\eta \langle 1|,  
\end{split}
\end{equation}
and subsequently, 
\begin{equation}
 W_n(\varphi,0) = (\sin\varphi)^n \1_{2^n}, \quad
  W_n^+(\varphi,0) = 0. 
\end{equation}
Setting 
\begin{align}
 &q_r(\varphi) = (\sin\varphi)^{-r+2} \sum_{\alpha_j \in \mathcal{J}} \langle 1| \bL^{\alpha_{2}}(\varphi,0) \cdots \bL^{\alpha_{r-1}}(\varphi,0) |1 \rangle
  \sigma^- \otimes \sigma^{\alpha_2} \otimes \cdots \otimes \sigma^{\alpha_{r-1}} \otimes \sigma^+, \\
 &p^+_{r}(\varphi) = (\sin\varphi)^{-r+2}
 \sum_{\alpha_j \in \{+,-,0,z\}} \langle 1| \bL^{\alpha_2}(\varphi,0) \cdots \bL^{\alpha_{r-1}}(\varphi,0) |1 \rangle
 \sigma^z \otimes \sigma^{\alpha_2} \otimes \cdots \otimes \sigma^{\alpha_{r-1}} \otimes \sigma^+, 
 \\
 &p^-_{r}(\varphi) = (\sin\varphi)^{-r+2}
 \sum_{\alpha_j \in \{+,-,0,z\}} \langle 1| \bL^{\alpha_2}(\varphi,0) \cdots \bL^{\alpha_{r-1}}(\varphi,0) |1 \rangle
 \sigma^- \otimes \sigma^{\alpha_2} \otimes \cdots \otimes \sigma^{\alpha_{r-1}} \otimes \sigma^z
\end{align}
leads to the simple expressions: 
\begin{align}
 &\sigma^- \otimes \partial_s W_{n-1}^+(\varphi,s)\Big|_{s=0}
  = 2\eta (\sin\varphi)^{n-2} \sum_{r=2}^n q_{r}(\varphi) \otimes \1_{2^{n-r}}, \\
 &\sigma^z \otimes \partial_s W_{n-1}^+(\varphi,s)\Big|_{s=0}
  = 2\eta (\sin\varphi)^{n-2} \sum_{r=2}^n p^+_{r}(\varphi) \otimes \1_{2^{n-r}}, \\
 &W_{n-1}^-(\varphi,0) \otimes \sigma^+ 
 = \sin\eta (\sin\varphi)^{n-2} \sum_{r=2}^n \1_{2^{n-r}} \otimes q_{r}(\varphi), \\
 &W_{n-1}^-(\varphi,0) \otimes \sigma^z 
 = \sin\eta (\sin\varphi)^{n-2} \sum_{r=2}^n \1_{2^{n-r}} \otimes p^-_{r}(\varphi). 
\end{align}
Note that the charge $Z_n(\varphi)$ \eqref{eq:quasilocal_c*} is also expressed in terms of $q_r(\varphi)$: 
\begin{equation} \label{eq:quasilocal_c}
 Z_n(\varphi) = \sum_{r=2}^n \sum_{k=0}^{n-k} \1_{2^{k}} \otimes q_r(\varphi) \otimes \1_{2^{n-r-k}}. 
\end{equation}
Then the commutator \eqref{eq:div_simple} is written by using $q_r(\varphi)$ and $p^{\pm}_r(\varphi)$ as 
\begin{equation}
\begin{split}
 [H,\,Z_n(\varphi)]
 &= \sigma^z \otimes \1_{2^{n-1}} - \1_{2^{n-1}} \otimes \sigma^z
 \\
 &+ (2\cot\varphi - \cot\xi_{\rm L}) \sin\eta \sum_{r=2}^n q_r(\varphi) \otimes \1_{2^{n-r}}
 - (2\cot\varphi - \cot\xi_{\rm R}) \sin\eta \sum_{r=2}^n \1_{2^{n-r}} \otimes q_r(\varphi)
 \\
 &+ \frac{\kappa_{\rm L} e^{\theta_{\rm L}}}{\sin\xi_{\rm L}} \sin\eta \sum_{r=2}^n p^+_r(\varphi) \otimes \1_{2^{n-r}}
 - \frac{\kappa_{\rm R} e^{-\theta_{\rm R}}}{\sin\xi_{\rm R}} \sin\eta \sum_{r=2}^n \1_{2^{n-r}} \otimes p^-_r(\varphi). 
\end{split}
\end{equation}

In order to show quasilocality of the charge $Z_n(\varphi)$, it is enough to show quasilocality of $q_r(\varphi)$ and $p^{\pm}_r(\varphi)$. We first compute the Hilbert-Schmidt norms of these quantities in the easy-plane anisotropy regime $\eta = \pi l/m$ for coprime $l,m \in \mathbb{Z}_{>0}, m \neq 0, l \leq m$. 
In this regime, each element of the Lax operators has the finite dimensional representation: 
\begin{equation}
\begin{split}
 &\bL^0(\varphi) = \sum_{k=0}^{m-1} \sin\varphi \cos\tfrac{\pi lk}{m} |k \rangle \langle k|, \quad
 \bL^z(\varphi) = -\sum_{k=1}^{m-1} \cos\varphi \sin\tfrac{\pi lk}{m} |k \rangle \langle k|, \\
 &\bL^+(\varphi) = -\sum_{k=1}^{m-2} \sin\tfrac{\pi lk}{m} |k+1 \rangle \langle k|, \quad
 \bL^-(\varphi) = \sum_{k=0}^{m-2} \sin\tfrac{\pi l(k+1)}{m} |k \rangle \langle k+1|. 
\end{split}
\end{equation}
This allows the explicit calculation of the Hilbert-Schmidt norms of $q_r(\varphi)$ and $p^{\pm}_r(\varphi)$ as 
\begin{equation} \label{eq:quasilocal2}
\begin{split}
 &(q_r(\varphi), q_r(\varphi)) \\
 &= \frac{1}{2^r} \sum_{\alpha_j,\alpha'_j \in \{+,-,0,z\}}
 \frac{1}{(\sin\varphi)^{r-2}} \Big( \langle 1| \bL^{\alpha'_2}(\varphi) \cdots \bL^{\alpha'_{r-1}}(\varphi) \Big)
 \frac{1}{(\sin\barphi)^{r-2}} \Big( \langle 1| \bL^{\alpha_2}(\barphi) \cdots \bL^{\alpha_{r-1}}(\barphi) \Big)^T \\
 &\hspace{5mm}\times
 \tr\Big( \sigma^+\sigma^- \otimes (\sigma^{\alpha_2})^T\sigma^{\alpha'_2} \otimes \cdots \otimes (\sigma^{\alpha_{r-1}})^T\sigma^{\alpha'_{r-1}} \otimes \sigma^-\sigma^+ \Big)
 \\
 &= \frac{1}{4} \langle 1|\bm{T}(\barphi,\varphi)^{r-2} |1 \rangle, 
\end{split}
\end{equation}
\begin{equation}
\begin{split}
 &(p^+_r(\varphi), p^+_r(\varphi)) \\
 &= \frac{1}{2^r} \sum_{\alpha_j,\alpha'_j \in \{+,-,0,z\}}
 \frac{1}{(\sin\varphi)^{r-2}} \Big( \langle 1| \bL^{\alpha'_2}(\varphi) \cdots \bL^{\alpha'_{r-1}}(\varphi) \Big)
 \frac{1}{(\sin\barphi)^{r-2}} \Big( \langle 1| \bL^{\alpha_2}(\barphi) \cdots \bL^{\alpha_{r-1}}(\barphi) \Big)^T \\
 &\hspace{5mm}\times
 \tr\Big( \sigma^z\sigma^z \otimes (\sigma^{\alpha_2})^T\sigma^{\alpha'_2} \otimes \cdots \otimes (\sigma^{\alpha_{r-1}})^T\sigma^{\alpha'_{r-1}} \otimes \sigma^-\sigma^+ \Big)
 \\
 &= \frac{1}{2} \langle 1|\bm{T}(\barphi,\varphi)^{r-2} |1 \rangle, 
\end{split}
\end{equation}
\begin{equation}
\begin{split}
 &(p^-_r(\varphi), p^-_r(\varphi)) \\
 &= \frac{1}{2^r} \sum_{\alpha_j,\alpha'_j \in \{+,-,0,z\}}
 \frac{1}{(\sin\varphi)^{r-2}} \Big( \langle 1| \bL^{\alpha'_2}(\varphi) \cdots \bL^{\alpha'_{r-1}}(\varphi) \Big)
 \frac{1}{(\sin\barphi)^{r-2}} \Big( \langle 1| \bL^{\alpha_2}(\barphi) \cdots \bL^{\alpha_{r-1}}(\barphi) \Big)^T \\
 &\hspace{5mm}\times
 \tr\Big( \sigma^+\sigma^- \otimes (\sigma^{\alpha_2})^T\sigma^{\alpha'_2} \otimes \cdots \otimes (\sigma^{\alpha_{r-1}})^T\sigma^{\alpha'_{r-1}} \otimes \sigma^z\sigma^z \Big)
 \\
 &= \frac{1}{2} \langle 1|\bm{T}(\barphi,\varphi)^{r-2} |1 \rangle. 
\end{split}
\end{equation}
Here we introduced 
\begin{equation}
 \bm{T}(\barphi,\varphi) = \sum_{k=1}^{m-1} 
  \Big( (\cos\tfrac{\pi lk}{m})^2 + \cot\barphi \cot\varphi (\sin\tfrac{\pi lk}{m})^2 \Big) |k \rangle \langle k|
  + \sum_{k=1}^{m-2} \frac{|\sin\tfrac{\pi lk}{m} \sin\tfrac{\pi l(k+1)}{m}|}{2\sin\barphi \sin\varphi} (|k \rangle \langle k+1| + |k+1 \rangle \langle k|). 
\end{equation}

Using the fact that the eigenvalues $\{\tau_i;\, i=1,\dots,m-1\}$ of $\bm{T}(\barphi,\varphi)$ satisfy $1 > |\tau_1| \geq \dots \geq |\tau_{m-1}|$, the Hilbert-Schmidt norms of $q_r(\varphi)$ and $p^{\pm}_r(\varphi)$ are subjected to 
\begin{equation}
 (q_r(\varphi), q_r(\varphi)) \leq \gamma e^{-\xi(\varphi) r}, \quad
  (p^{\pm}_r(\varphi), p^{\pm}_r(\varphi)) \leq \gamma^{\pm} e^{-\xi(\varphi) r}
\end{equation}
with the decay constant: 
\begin{equation}
 \xi(\varphi) = -\frac{1}{2} \log \tau_1(\varphi) > 0. 
\end{equation}
Therefore, $Z_n(\varphi)$ is quasilocal and bounded by 
\begin{equation}
\begin{split}
 (Z_n(\varphi), Z_n(\varphi)) &= n \sum_{r=2}^n \Big( 1 - \frac{r-1}{n} \Big) (q_r(\varphi), q_r(\varphi)) 
 \leq n \gamma^2 \sum_{r=2}^n e^{-2\xi(\varphi) r}
 < n \frac{\gamma^2}{1 - e^{-2\xi(\varphi)}}. 
\end{split}
\end{equation}

Besides the quasilocal charges thus constructed, the normal local charges $Q_r$ may also contribute to the lower bound, since, unlike the periodic or trivial open boundary cases, the open XXZ chain with \remove{arbitrary}\add{non-diagonal} boundary magnetic fields does not \add{in general} have the spin-reversal symmetry and, subsequently, the conventional local charges include odd parity part as well as the even parity part. In the case where 
\begin{equation} \label{eq:parity_inv}
 \xi_{\rm L,R} = \frac{\pi}{2}, \quad
  \theta_{\rm L,R} = i\pi n, \quad n \in \mathbb{Z}, 
\end{equation}
the local charges consist only of even parity part and thus the normal local charges have no contribution to the lower bound. 

\subsection{Discussion}
\subsubsection{The open XXZ spin chain with spin-reversal symmetry}
When the spin chain possesses the spin-reversal symmetry realized by the choice of parameter values \eqref{eq:parity_inv}, the optimized Mazur bound is solely evaluated through the formula \eqref{eq:Mazur_b} by using the quasilocal charges constructed in the previous section \eqref{eq:quasilocal_c}. 
Noting that the spin current is given by 
\begin{equation}
 J = i \sum_{j=1}^{n-1}(\sigma_j^+ \sigma_{j+1}^- - \sigma_j^- \sigma_{j+1}^+), 
\end{equation}
the overlap between the current and the quasilocal charge has a constant value $Z_J(\varphi) = i/4$. On the other hand, the inner product of the quasilocal charges is computed by using \eqref{eq:quasilocal2}: 
\begin{equation}
\begin{split}
 (Z_n(\varphi), Z_n(\varphi')) 
 &= \sum_{r=2}^n (n-r+1) \tr \Big( (q_r(\varphi))^{\dag} q_r(\varphi) \Big) \\
 &= \frac{n}{4} \sum_{r=2}^{\infty} \langle 1| \bT(\varphi,\varphi')^{r-2} |1 \rangle + \mathcal{O}(1) \\
 &= \frac{n}{4} \langle 1| (\1 - \bT(\varphi,\varphi'))^{-1} |1 \rangle + \mathcal{O}(1) \\
 &= -n \frac{\sin\varphi \sin\varphi'}{2 \sin^2 \frac{\pi l}{m}} \frac{\sin((n-1)(\varphi+\varphi'))}{\sin(m(\varphi+\varphi'))} + \mathcal{O}(1). 
\end{split}
\end{equation}
Therefore, the complex Fredholm equation of the first kind \eqref{eq:variation} is solved as 
\begin{equation}
 f(\varphi) = -\frac{im}{\pi} \frac{\sin^2\frac{\pi l}{m}}{|\sin\varphi|^4}. 
\end{equation}
Finally, the lower bound on the Drude weight is evaluated as 
\begin{equation}
 D_Z \geq \frac{1}{4} \frac{\sin^2\frac{\pi l}{m}}{\sin^2\frac{\pi}{m}} \Big( 1 - \frac{m}{2\pi} \sin\Big( \frac{2\pi}{m} \Big) \Big), 
\end{equation}
which coincides with the bound for the periodic and trivial open boundary cases. 

\subsubsection{The open XXZ spin chain without spin-reversal symmetry}
The open XXZ chain with \remove{arbitrary}\add{non-diagonal} boundary magnetic fields in general does not have the spin-reversal symmetry and, consequently, its local charges contain both even and odd parity parts. 
By taking into account of the contributions of the local charges, the Mazur bound is expected to be further optimized. 

Let us consider the operator $B$ consisting of both local and quasilocal charges: 
\begin{equation}
 B = \overline{J} - \int_{\mathcal{D}_m} d^2\varphi\, f(\varphi) Z(\varphi) - \sum_{r=1}^n \alpha_r Q_r, 
\end{equation}
where $\overline{J}$ is the time-averaged current: 
\begin{equation}
 \overline{J} := \lim_{T \to \infty} \frac{1}{T} \int_0^T dt\, e^{iHt} J e^{-iHt}. 
\end{equation}
Note that the local charges are defined by the logarithmic derivatives of the transfer matrix: 
\begin{equation}
\begin{split}
 &Q_r = \partial^{r-1}_{\varphi} \ln V_n(\varphi,\tfrac{1}{2}) \Big|_{\varphi=\frac{\eta}{2}}, \\
 &V_n(\varphi,\tfrac{1}{2}) = \Big(\frac{-1}{\sinh(\varphi-\frac{\eta}{2}) \sinh(\varphi+\frac{3\eta}{2})}\Big)^n \\
 &\times \textstyle\tr_a \Big( K(\varphi;\xi_{\rm L},\kappa_{\rm L},\theta_{\rm L}) \bL_1(\varphi,\tfrac{1}{2}) \dots \bL_n(\varphi,\tfrac{1}{2}) K(\varphi+\eta;\xi_{\rm R},\kappa_{\rm R},\theta_{\rm R}) \bL_1(\varphi+\eta,\tfrac{1}{2}) \dots \bL_n(\varphi+\eta,\tfrac{1}{2}) \Big), 
\end{split}
\end{equation}
where we used the relation 
\begin{equation}
 \bL^{-1}(\varphi,\tfrac{1}{2}) = \frac{-1}{\sinh(\varphi+\frac{\eta}{2}) \sinh(\varphi-\frac{3\eta}{2})} \bL(-\varphi+\eta,\tfrac{1}{2}). 
\end{equation}
The local charges are expressed by the corresponding local densities $q^{(r)}$: 
\begin{equation}
 Q_r = \sum_{x=0}^{n-1} (\1_{2^{x}} \otimes q^{(r)} \otimes \1_{2^{n-r-x}}). 
\end{equation}

Since the following inequality relation holds for the Hilbert-Schmidt inner product of $B$: 
\begin{equation} \label{eq:Mazur_b**}
\begin{split}
 0 \leq \frac{1}{2n}(B,B)
 &= D_{ZQ} - \frac{1}{2n} \int_{\mathcal{D}_m} d^2\varphi\, f(\varphi) (J,Z(\varphi)) - \frac{1}{2n} \int_{\mathcal{D}_m} d^2\varphi\, \overline{f(\varphi)} (Z(\varphi),J) \\
 &- \frac{1}{2n} \sum_{r=0}^n \alpha_r (J,Q_r) - \frac{1}{2n} \sum_{r=0}^n \overline{\alpha_r} (Q_r,J) \\
 &+ \frac{1}{2n} \sum_{r=0}^n \overline{\alpha_r} \int_{\mathcal{D}_m} d^2\varphi\, f(\varphi) (Q_r,Z(\varphi)) + \frac{1}{2n} \sum_{r=0}^n \alpha_r \int_{\mathcal{D}_m} d^2\varphi\, \overline{f(\varphi)} (Z(\varphi),Q_r) \\
 &+ \frac{1}{2n} \int_{\mathcal{D}_m} d^2\varphi \int_{\mathcal{D}_m} d^2\varphi'\, \overline{f(\varphi)} f(\varphi') (Z(\varphi),Z(\varphi')) + \frac{1}{2n} \sum_{r,r'=0}^n \overline{\alpha_r}\alpha_{r'} (Q_r,Q_{r'}), 
\end{split}
\end{equation}
the high-temperature Drude weight is bounded from below by 
\begin{equation}
\begin{split}
 D_{ZQ} &\geq F_n[f,\{\alpha_r\}] \\
 &:= \int_{\mathcal{D}_m} d^2\varphi\, {\rm Re} \Big( \frac{1}{n}(J,Z(\varphi)) f(\varphi) \Big)
 + \sum_{r=1}^{n} {\rm Re} \Big( \frac{1}{n} (J,Q_r) \alpha_r \Big) \\
 &- \sum_{r=1}^{n} {\rm Re} \Big( \alpha_r \int_{\mathcal{D}_m} d^2\varphi\, \frac{1}{n}(Q_r,Z(\varphi)) f(\varphi) \Big) \\
 &- \int_{\mathcal{D}_m} d^2\varphi \int_{\mathcal{D}_m} d^2\varphi'\, \frac{1}{2n}(Z(\varphi),Z(\varphi')) \overline{f(\varphi)} f(\varphi') 
 - \sum_{r,r'=1}^{n} \frac{1}{2n}(Q_r,Q_{r'}) \overline{\alpha_r} \alpha_{r'}. 
\end{split}
\end{equation}
The function $f(\varphi)$ and the parameter $\alpha_r$ are determined through the variation and differentiation of $F_n[f,\{\alpha_k\}]$: 
\begin{equation} \label{eq:functional*}
\begin{split}
 &\delta F_n[f] = {\rm Re} \int_{\mathcal{D}_m} d^2\varphi\, \overline{\delta f(\varphi)} \Big( \frac{1}{n}\overline{(J,Z(\varphi))} - \sum_{r=1}^{n} \frac{1}{n} \overline{(Q_r,Z(\varphi))}\, \overline{\alpha_r}  - \int_{\mathcal{D}_m} d^2\varphi'\, \frac{1}{n} (Z(\varphi),Z(\varphi')) f(\varphi') \Big) 
 =0, \\
 &\frac{d F_n[\alpha_r]}{d \alpha_r} = {\rm Re} \Big( \frac{1}{n}\overline{(J,Q_r)} - \int_{\mathcal{D}_m} d^2\varphi\, \frac{1}{n} \overline{(Q_r,Z(\varphi))}\, \overline{f(\varphi)}  - \sum_{r'=1}^{n} \frac{1}{n} (Q_r,Q_{r'}) \alpha_{r'}  \Big)=0. 
\end{split}
\end{equation}
Substituting \eqref{eq:functional*} into \eqref{eq:Mazur_b**}\remove{ and taking the $n \to \infty$ limit}, we find 
\begin{equation}
\begin{split}
 &D_{ZQ} \geq \frac{1}{2} \int_{\mathcal{D}_m} d^2\varphi\, {\rm Re} \Big( Z_J(\varphi)f(\varphi) \Big)
 + \frac{1}{2} \sum_{r=1}^{n} {\rm Re} \Big( Q_{r,J} \alpha_r \Big)
 = D_Z, \\
 &Q_{r,J} := \lim_{n \to \infty} \frac{1}{n} (J, Q_r). 
\end{split}
\end{equation}
\add{in the $n \to \infty$ limit.} The first term is what was evaluated in the previous section. \add{Although $Q_{r,J} \neq 0$ in the spin chain without spin-reversal symmetry, the second term vanishes in the thermodynamic limit, since the boundary effect on the local charges localizes at the outermost sites of the spin chain.}\remove{The second inequality holds since $Q_{r,J} \neq 0$ in the spin chain without spin-reversal symmetry}. \comment{IS THIS REALLY TRUE, NAMELY BREAKING OF SPIN REVERSAL SYMMETRY IS ONLY BY THE BOUNDARY TEMRS, SO ITS EFFECT WILL PROBABLY VANISH IN THE THERMODYNAMIC LIMIT?}
\section{Conclusion}
In this paper, we have showed that for a certain family of the spin-$1/2$ XXZ chain the exact NESS density operator is derived \remtomaz{under}\addtomaz{in} the \remtomaz{existence}\addtomaz{presence} of\remtomaz{arbitrary} boundary magnetic fields. By properly choosing the dissipation rates as functions of boundary parameters, we found that the NESS density operator is given by the NHTO in spite of \remtomaz{arbitrary choice}\addtomaz{a choice of diagonal and non-diagonal boundary fields}. 
We have also derived the quasilocal charges of the corresponding spin chain without boundary dissipation. We found that the NHTO satisfies quasilocality even in the system without spin-reversal symmetry. In such a system, the conventional local charges possess the odd parity part besides the even parity part. As a consequence, we showed \remove{the existence of the further optimized Mazur bound by using the non-orthogonality of the current and the local charges. This is not due to the boundary effect on the bulk behavior but simply due to the symmetry of the Hamiltonian.}\add{that the terms coming from the odd parity part of the local charges have non-zero values, although they vanish in the thermodynamic limit. 
We \remtomaz{shall mention}\addtomaz{note} that the local charges also give non-zero contribution in the spin chain with the bulk \addtomaz{homogeneous} magnetic field, although their effect vanishes in the high temperature limit. }

Although we discussed the nonequilibrium behavior of the integrable spin chain with boundaries, we did not find any perception in the context of the reflection relation~\cite{bib:S88}. Furthermore, we did not \remtomaz{investigate the}\addtomaz{find any integrable} NESS density operators which are not expressed \addtomaz{in terms of} the NHTO. \remove{By using the inhomogeneous rotational transformation in the $xy$-plane, the homogeneous version of which we used in this paper to obtain the $U(1)$-symmetry of the Hamiltonian, we expect that the NESS density operator is modified in such a way that includes the inhomogeneity parameters.}\add{We leave these questions as future works.} \comment{THIS LAST SENTENCE I DON'T UNDERSTAND.}

\section*{Acknowledgements}
C. M. is supported by JSPS Grant-in-Aid, No. 15K20939, Japan and JST CREST, No. JPMJCR14D2, Japan.  T. P. is supported by Grants P1-0044, N1-0025 and N1-0055 of Slovenian Research Agency, and ERC Grant OMNES.

\bibliographystyle{abbrv}
\bibliography{references}

\end{document}